\documentstyle[10pt,aaspp4]{article}
\begin{document}
\title{A Cubic Micron of Equilibrium Pair Plasma?}
\author{J. I. Katz}
\affil{Department of Physics and McDonnell Center for the Space Sciences,
Washington University, St. Louis, Mo. 63130}
\authoremail{katz@wuphys.wustl.edu}
\begin{abstract}
Is it possible to create a small volume of equilibrium pair plasma in the
laboratory with an intense laser pulse?  No.  The required power would 
exceed $10^{18}$ W.  The number of seed particles required to absorb the laser
energy would approach the number of pairs in the desired plasma, and their
radiation in a visible laser field would occur at excessively high frequencies.
\end{abstract}
\section{Introduction}
Space in thermal equilibrium at sufficiently high temperature is filled with
an electron-positron pair plasma.  If there are no baryons present the
equilibrium densities of electrons and positrons are equal and are given by
$$n_\pm = {1 \over \pi^2 \hbar^3} \int^\infty_0 {p^2\, dp \over \exp{(p^2 +
m^2 c^4)^{1/2}/k_B T} + 1}, \eqno(1)$$
where $m$ is the electron's mass and other symbols have their usual meanings
(\cite{LL58}).  If $k_B T \ll m c^2$ this result is approximated
$$n_\pm \approx 2 \left({m k_B T \over 2 \pi \hbar^2}\right)^{3/2} \exp{(- m
c^2/k_B T}). \eqno(2)$$
A pair plasma with this density is in complete thermodynamic equilibrium
with the photon gas into which it annihilates and from which it is formed by
pair production, and each species has zero chemical potential.  The
radiation field has a Planckian spectrum at temperature $T$.

It is necessary to distinguish this pair plasma in equilibrium with a black
body radiation field from the much lower density pair plasmas frequently
considered in astrophysics.  These low density pair plasmas may have
equilibrium relativistic Maxwellian particle distribution functions, but
their densities are lower, usually by many orders of magnitude, than would
be given by Eq. 1.  Their chemical potentials are negative and many times $m
c^2$ in magnitude.  They are not in equilibrium with a Planckian radiation
field.  Their optical depth is generally very low, and the radiation energy
density is much less than that of the pairs.  Such low density pair plasmas
have been considered by many authors in many astrophysical contexts, but
will not be discussed further here.

Completely equilibrium pair plasmas occur (with the addition of baryons and
a consequent shift in chemical potential and inequality between $n_-$ and
$n_+$) in the interior of very evolved stars, where they are hidden from
view.  Equilibrium pair plasmas have also been considered (\cite{TD95,K96})
as sources of emission from soft gamma repeaters (SGR), if trapped within
the strong magnetic field of a neutron star.  The minimum required magnetic
field is only $\sim 10^{11}$ gauss, although much stronger fields have been
considered.  Directly observed pair plasma would be expected to produce an
approximately black body radiation spectrum, whose temperature is only
weakly dependent on the many unknown astrophysical parameters.  This is in
at least qualitative agreement with observations of SGR, whose spectra
appear to be strongly self-absorbed at low frequencies, to be characterized
by temperatures roughly comparable to those predicted by pair plasma models,
and which are nearly uniform from burst to burst and within bursts.  In
addition, an equilibrium pair plasma is such a simple state of matter, which
occurs whenever the energy density is sufficient and the optical depth
large, that it may be found in unanticipated places in the universe and is
likely to repay study.
\section{Equilibrium Pair Plasmas in the Laboratory?}
\subsection{Energetics}
The advent of high power lasers, now including the Petawatt ($10^{15}$ W)
laser, raises the question of whether it might be possible to create a
small volume of equilibrium pair plasma in the laboratory.  The first
criterion which must be satisfied is energetic:  Can a laser supply the
power lost from the surface of the plasma?

There are two mechanisms of energy loss.  The surface will radiate
approximately as a black body at its temperature, and the pairs themselves
will expand freely (in contrast to the situation near a neutron star, the
magnetic field required to confine the pairs is unachievable in the
laboratory).  It is easy to see that the second process is less important
than the first, partly because the pairs escape at a slower speed than
photons, and partly because the equilibrium pair energy density is always 
less than 7/4 that of photons at the same temperature, and is less by an
exponentially large factor if $k_B T \ll m c^2$, the case of interest, 
because of the exponential in Eq. 2.  We will therefore only consider energy
loss by radiation.

The minimum temperature required to obtain an equilibrium pair plasma may be
estimated by requiring that the Thomson scattering optical depth be at least
3, in order that photon-pair equilibrium be achieved.  Assume a spherical
plasma of radius $r = 0.6\,\mu$, which has a volume of nearly a cubic
micron.  This size is chosen because it is approximately the size of a
diffraction-limited focal spot of a visible laser.  Then the required pair
density, allowing for the presence of both species, is
$$n_\pm = {3 \over 2 \sigma_{es} r} \approx 4 \times 10^{28}\ {\rm cm}^{-3}.
\eqno(3)$$
This is probably an underestimate of the actual required density because of
the decrease of cross-section with increasing energy and the incomplete
equilibration achieved in Compton scattering in contrast to absorption.

The required temperature may be found from Eqs. 3 and 4, with the result $T
= 2.3 \times 10^{9\ \circ}$K and the parameter
$$u \equiv {k_B T \over m c^2} = 0.39. \eqno(4)$$
The resulting black body intensity radiated
$$I = {2 \pi^5 \over 15} {m^4 c^6 \over h^3} u^4 = 1.6 \times 10^{26}\ {\rm W
/ cm}^2. \eqno(5)$$
The total power radiated
$$P = 4 \pi r^2 I = 7 \times 10^{18}\ {\rm W}. \eqno(6)$$
This power need be radiated only for a time $O(r/c) \sim 2 \times 10^{-15}\,
{\rm s}$, so the required energy is only $\sim 15$ kJ.  However, such brief
illumination would require a laser of a bandwidth far broader than any known
today.

These numbers are intimidating.  The power exceeds that of the Petawatt
laser by nearly four orders of magnitude.  While this would not violate any
fundamental physical bound, it is certainly beyond anything achievable
within forseeable budgets and with forseeable technology.

Larger volumes of pair plasma would require even more power.  First suppose
$n_\pm \propto T^3$, as is the case for $k_B T \gg m c^2$.  Then the
temperature required (by the condition that the optical depth equal 3) $T
\propto n_\pm^{1/3} \propto r^{-1/3}$.  The radiated power $P \propto T^4 r^2
\propto r^{2/3}$ increases as $r$ increases.  In fact, in the applicable
regime $k_B T \ll m c^2$ the decrease of $T$ with decreasing $n_\pm$ and 
increasing $r$ is slower (the sensitivity of $n_\pm$ to $T$ is greater, 
because of the exponential in Eq. 2), and the dependence of $P$ on $r$ 
approaches the second power.  It is clear that increasing $r$ only increases
the required power.

Values of $r$ significantly less than 0.6 $\mu$ are excluded, so long as
power is provided by a laser of near-visible wavelength, by the diffraction
limit on focal spot size.  However, even if much smaller $r$ could be 
obtained the required power would not be much less.  The reason for this is 
that values of $r$ less than a few tenths of a micron imply $k_B T > m c^2$,
and the Thomson cross-section must be replaced by the Klein-Nishina 
cross-section, which decreases approximately $\propto T^{-1}$.  Taking 
$n_\pm \propto T^3$ yields $r \propto T^{-2}$ and $P \propto T^4 r^2$, 
independent of $T$ and $r$.
\subsection{Kinetics}
Suppose that, somehow, visible lasers of the required power were available.
The implied electric and magnetic fields would be impressive, with amplitudes
$$B_0 = E_0 = \left({8 \pi I \over c}\right)^{1/2} = \left({16 \pi^6 \over
15}\right)^{1/2} {m^2 c^{5/2} \over h^{3/2}} u^2 = 1.2 \times 10^{12} 
\left({u \over 0.39}\right)^2\quad{\rm(cgs\ units)}. \eqno(7)$$
These should be compared to the characteristic quantum fields at which Landau
levels are spaced by $m c^2$ and spontaneous electric pair production is
rapid $B_q = E_q \equiv m^2 c^3 / e \hbar$:
$$v \equiv {E_0 \over E_q} = \left({2 \pi^3 \over 15} \alpha\right)^{1/2} 
u^2 = 0.026 \left({u \over 0.39}\right)^2, \eqno(8)$$
where $\alpha \equiv e^2 / \hbar c$ is the usual fine-structure constant.

We now ignore the possibility of spontaneous pair production, and consider
only the effects of such large fields on an injected electron or a small
number of pairs.  Will their motion lead to rapid radiation of the soft
gamma-rays which can create the desired equilibrium pair plasma?  We
consider only the radiation by individual particles in intense fields, and
also ignore such processes as inverse bremsstrahlung, which are generally 
slow at high energies and low densities.

The theory of the motion of a charged particle in a strong electromagnetic
wave was recently reviewed by \cite{M97}, and many useful results are
presented by \cite{GO71}.  I will apply results which assume a plane
space-filling electromagnetic wave.  In fact, these results overestimate by
a very large factor the energy absorbed in a small laser focal spot by 
charged particles which soon leave the region of high intensity.  

The intensity of an electromagnetic field of angular frequency $\omega$ and 
wavelength $\lambda = 2 \pi c / \omega$ is described by a parameter
$$\eta \equiv {e E_0 \over m c \omega} = 5500 \left({u \over 0.39}\right)^2
{\lambda \over 0.5\,\mu}. \eqno(9)$$
Radiation reaction is described by a parameter
$$\epsilon \equiv {e^2 \omega \eta^2 \over 3 m c^3} = 0.35 \left({u \over
0.39}\right)^4 {\lambda \over 0.5\,\mu}. \eqno(10)$$

The radiation parameter is much larger than that found for pulsar fields, 
for which $\epsilon \ll 1$, but is small enough that the results of 
\cite{GO71} are still approximately valid.  The mean power radiated per
electron (or positron) in the laser field is given by \cite{GO71}, in a
process which they name nonlinear inverse Compton scattering,
$$P_r = {e^4 E_0^2 \over 3 m^2 c^3} = 110 \left({u \over 0.39}\right)^4\
{\rm W}. \eqno(11)$$
Approximately $6 \times 10^{16}$ electrons and positrons must be present to
absorb the power required (Eq. 6).  This is approximately equal to the $8
\times 10^{16}$ particles present in a cubic micron of equilibrium pair
plasma (Eq. 3).  In other words, the laser could sustain the plasma once it
was created, but could not produce it from a much smaller number of seed
particles by nonlinear inverse Compton scattering.

\cite{GO71} also estimate the critical frequency $\nu_{crit}$ of nonlinear
inverse Compton scattering for a very energetic particle (Lorentz factor
$\gg \eta m c^2$) entering the laser field.  Their argument may also be
applied to particles accelerated by the laser field itself, with similar
results:
$$\nu_{crit} \approx \omega \eta^3 \approx {v^3 \over w^2} {m c^2 \over
\hbar} \approx 10^{26} \left({u \over 0.39}\right)^6 \left({\lambda \over 
0.5\,\mu}\right)^2\ {\rm Hz}, \eqno(12)$$
where the parameter $w \equiv \hbar \omega / m c^2$.

These are photons of $\sim 10^{11}$ eV energy.  Their cross-section for
photon-photon pair production is very small, and they are produced (because
of their high energy) in very small numbers even if sufficient
charged particles were present to absorb the required power.  If produced
they will escape, rather than producing an equilibrium pair plasma.
\section{Discussion}
It is apparent that near-visible lasers cannot be used to produce an
equilibrium pair plasma, even if sufficient power were available.  This goal
requires not only extremely high power (Eq. 6) but also that this power be 
delivered in photons of energy approaching $m c^2$ (Eq. 12).  Until
gamma-ray lasers are developed, this is likely to remain unachievable.
\acknowledgements
I thank D. D. Meyerhofer for useful discussions.

\end{document}